\input harvmac.tex
\input diagrams

\let\includefigures=\iftrue
%
\let\useblackboard=\iftrue
%
%
\newfam\black
\input epsf
\includefigures
\message{If you do not have epsf.tex (to include figures),}
\message{change the option at the top of the tex file.}
\def\figin{\epsfcheck\figin}\def\figins{\epsfcheck\figins}
\def\epsfcheck{\ifx\epsfbox\UnDeFiNeD
\message{(NO epsf.tex, FIGURES WILL BE IGNORED)}
\gdef\figin##1{\vskip2in}\gdef\figins##1{\hskip.5in}
\else\message{(FIGURES WILL BE INCLUDED)}%
\gdef\figin##1{##1}\gdef\figins##1{##1}\fi}
\def\DefWarn#1{}
\def\figinsert{\goodbreak\midinsert}
\def\ifig#1#2#3{\DefWarn#1\xdef#1{fig.~\the\figno}
\writedef{#1\leftbracket fig.\noexpand~\the\figno}%
\figinsert\figin{\centerline{#3}}\medskip\centerline{\vbox{\baselineskip12pt
\advance\hsize by -1truein\noindent\footnotefont{\bf Fig.~\the\figno:} #2}}
\bigskip\endinsert\global\advance\figno by1}
\else
\def\ifig#1#2#3{\xdef#1{fig.~\the\figno}
\writedef{#1\leftbracket fig.\noexpand~\the\figno}%
\global\advance\figno by1}
\fi

\def\IZ{\relax\ifmmode\mathchoice
{\hbox{\cmss Z\kern-.4em Z}}{\hbox{\cmss Z\kern-.4em Z}}
{\lower.9pt\hbox{\cmsss Z\kern-.4em Z}}
{\lower1.2pt\hbox{\cmsss Z\kern-.4em Z}}\else{\cmss Z\kern-.4em
Z}\fi}
\def\inbar{\,\vrule height1.5ex width.4pt depth0pt}
\def\IB{\relax{\rm I\kern-.18em B}}
\def\IC{\relax\hbox{$\inbar\kern-.3em{\rm C}$}}
\def\ID{\relax{\rm I\kern-.18em D}}
\def\IE{\relax{\rm I\kern-.18em E}}
\def\IF{\relax{\rm I\kern-.18em F}}
\def\IG{\relax\hbox{$\inbar\kern-.3em{\rm G}$}}
\def\IH{\relax{\rm I\kern-.18em H}}
\def\II{\relax{\rm I\kern-.18em I}}
\def\IK{\relax{\rm I\kern-.18em K}}
\def\IL{\relax{\rm I\kern-.18em L}}
\def\IM{\relax{\rm I\kern-.18em M}}
\def\IN{\relax{\rm I\kern-.18em N}}
\def\IO{\relax\hbox{$\inbar\kern-.3em{\rm O}$}}
\def\IP{\relax{\rm I\kern-.18em P}}
\def\IQ{\relax\hbox{$\inbar\kern-.3em{\rm Q}$}}
\def\IR{\relax{\rm I\kern-.18em R}}
\font\cmss=cmss10 \font\cmsss=cmss10 at 7pt
\def\IZ{\relax\ifmmode\mathchoice
{\hbox{\cmss Z\kern-.4em Z}}{\hbox{\cmss Z\kern-.4em Z}}
{\lower.9pt\hbox{\cmsss Z\kern-.4em Z}}
{\lower1.2pt\hbox{\cmsss Z\kern-.4em Z}}\else{\cmss Z\kern-.4em Z}\fi}
\def\IGa{\relax\hbox{${\rm I}\kern-.18em\Gamma$}}
\def\IPi{\relax\hbox{${\rm I}\kern-.18em\Pi$}}
\def\ITh{\relax\hbox{$\inbar\kern-.3em\Theta$}}
\def\IOm{\relax\hbox{$\inbar\kern-3.00pt\Omega$}}

\def\inbar{\,\vrule height1.5ex width.4pt depth0pt}

\font\cmss=cmss10 \font\cmsss=cmss10 at 7pt
\def\IR{\relax{\rm I\kern-.18em R}}

\lref\dtopics{M.R. Douglas ``Topics in D-geometry'',  Class. Quant. Grav.
{\bf 17} (2000) 1057, hep-th/9910170.}
\lref\dbranes{M.R. Douglas ``D-branes on Calabi Yau Manifolds'', 
math.AG/0009209.}
\lref\dd{D.E. Diaconescu, M.R. Douglas ``D-branes on stringy Calabi Yau 
manifolds'', hep-th/0006224.}
\lref\lerche{W. Lerche, ``On a boundary CFT description of nonperturbative
${\cal N}=2$ Yang-Mills theory'', hep-th/0006100.}
\lref\fm{B. Fiol and M. Mari\~no, ``BPS states and algebras from quivers'',
JHEP {\bf 0007} (2000), 31.}
\lref\hm{J. Harvey and G. Moore, ``On the algebras of BPS states'',
Commun. Math. Phys.{\bf 197} (1998) 489.}
\lref\pistab{M.R. Douglas, B. Fiol and C. R\"omelsberger ``Stability and BPS 
branes'', hep-th/0002037.}
\lref\index{M.R. Douglas and B. Fiol, ``D-branes and discrete torsion II'',
hep-th/9903031.}
\lref\noncompact{M.R Douglas, B. Fiol and C. R\"omelsberger ``The spectrum 
of BPS branes on a non-compact Calabi-Yau'', hep-th/0003263.}
\lref\geoeng{S. Katz, A. Klemm and C. Vafa, ``Geometric engineering of quantum
field theories'', Nucl. Phys. {\bf B  497}(1997), 173.}
\lref\dm{M.R. Douglas and G. Moore, ``D-branes, ALE spaces and quivers'', 
hep-th/9603167.}
\lref\shamit{S. Kachru and C. Vafa ``Exact Results for ${\cal N}=2$ 
Compactifications of Heterotic Strings'', Nucl. Phys. {\bf B 450} (1995) 69. 
S.~Kachru, A.~Klemm, W.~Lerche, P.~Mayr and C.~Vafa, ``Nonperturbative 
results on the point particle limit of N=2 heterotic string 
compactifications,'' Nucl. Phys. {\bf B459} (1996), 537.}
\lref\selfdual{A. Klemm, W. Lerche, P. Mayr, C. Vafa and N. Warner, ``
Self-dual strings and ${\cal N}=2$ supersymmetric field theory'', Nucl.Phys.
{\bf B477} (1996) 746.}
\lref\mirror{S. Katz, P. Mayr and C. Vafa ``Mirror symmetry and exact solution
of 4d ${\cal N}=2$ gauge theories. I'', Adv. Theor. Math. Phys. 
{\bf 1} (1998) 53.}
\lref\dg{D.E. Diaconescu and J. Gomis, ``Boundary states and fractional 
branes'', JHEP {\bf 0010} (2000), 001.}
\lref\witten{E. Witten, ``Phases of ${\cal N}=2$ theories in two-dimensions'',
Nucl.Phys. {\bf B403} (1993) 159.}
\lref\govinda{S. Govindarajan and T. Jayaraman, ``D-branes, exceptional
sheaves and quivers on Calabi-Yau manifolds: from Mukai to McKay'', 
hep-th/0010196.}
\lref\tomasi{A. Tomasiello, ``D-branes on Calabi-Yau manifolds and helices.'',
hep-th/0010217.}
\lref\pmayr{P. Mayr, ``Phases of supersymmetryc D-branes on K\"ahler 
manifolds and the McKay correspondence'', hep-th/0010223.}
\lref\king{A.D. King, ``Moduli of representations of finite 
dimensional algebras,'' Quart. J. Math. Oxford {\bf 45} (1994) 515. }
\lref\katz{S. Katz, ``Toric Geometry for String Theory'', Lecture Notes for
the CMI Summer School on Mirror Symmetry, June 2000.}
\lref\lerch{W. Lerche, A. Lutken and C. Schweigert, ``D-branes on ALE Spaces
and the ADE Classification of Conformal Field Theories'',hep-th/0006247.} 
\lref\zaslow{E. Zaslow ``Topological Orbifold Models and Quantum Cohomology 
Rings'', Commun. Math. Phys. {\bf 156} (1993) 301.}
\lref\quintic{I. Brunner, M.R. Douglas, A. Lawrence and C. R\"omelsberger,
``D-branes on the Quintic''. JHEP {\bf 0008} (2000), 015.}
\lref\dgm{M.R. Douglas, B. Greene and D. Morrison ``Orbifold Resolution by
D-Branes''Nucl.\ Phys.\  {\bf B506} (1997), 84.}
\lref\mayr{P. Mayr ``Geometric Construction of ${\cal N}=2$ gauge 
theories'', Fortsch. Phys. {\bf 47} (1999), 39. hep-th/9807096.}
\lref\lercherev{W. Lerche 
``Introduction to Seiberg-Witten theory and its stringy origin''. 
Nucl. Phys. Proc. Suppl. {\bf 55B} (1997), 83. Fortsch. Phys. {\bf 45} (1997)
 293.}
\lref\klemrev{A. Klemm ``On the geometry behind ${\cal N}=2$ 
supersymmetric effective actions in four-dimensions'', hep-th/9705131.}
\lref\kac{V.G. Kac, ``Infinite root systems, representations of 
graphs and invariant theory,'' Inv. Math. {\bf 56} (1980) 57.}
\lref\bife{A. Bilal, F. Ferrari ``The Strong-Coupling Spectrum of the 
Seiberg-Witten Theory'',Nucl.Phys. {\bf B 469} (1996), 387. 
`Curves of Marginal Stability and Weak and Strong-Coupling BPS
Spectra in $N=2$ Supersymmetric QCD'', Nucl.Phys. {\bf B 480} (1996) 589}
\lref\scho{A. Schofield, ``Birational classification of moduli spaces of
representations of quivers'', math.AG/9911014.}
\lref\schotwo{A. Schofield, ``General representations of quivers'', Proc.
London Math. Soc. (3) {\bf 65} (1992) no.1, 46.}
\lref\sw{N. Seiberg, E. Witten, ``Monopole Condensation, And Confinement In 
${\cal N=2}$ Supersymmetric Yang-Mills Theory'', Nucl.Phys. {\bf B 426} 
(1994) 19; Erratum-ibid. {\bf B 430} (1994) 485.}
\lref\cate{M.R. Douglas, ``D-branes, Categories and ${\cal N}$=1 
Supersymmetry'', hep-th/0011017.}
\lref\kmp{S. Katz, D.R. Morrison and R. Plesser, ``Enhanced Gauge Symmetry
in type II'', Nucl. Phys. {\bf B 477} (1996), 105.}
\lref\iqbal{A. Iqbal, K. Hori and C. Vafa, ``D-Branes and Mirror 
Symmetry'', hep-th/0005247.}
\lref\km{A. Klemm and P. Mayr ``Strong Coupling Singularities and 
Non-abelian Gauge Symmetries in $N=2$ String Theory'' Nucl. Phys. 
{\bf B 469} (1996), 37.}
\lref\agm{P. Aspinwall, B. Greene and D. Morrison ``Measuring Small Distances 
in N=2 Sigma Models'', Nucl. Phys. {\bf B 420} (1994) 184.}
\lref\hollow{C. Fraser and T. Hollowod, ``On the weak coupling spectrum of 
${\cal N=2}$ supersymmetric SU(n) gauge theory''. Nucl.Phys. {\bf B 490} 
(1997), 217.}
\lref\index{M. Berkooz and M.R. Douglas ``Five-branes in M(atrix) Theory'',
Phys.Lett. {\bf B 395} (1997) 196. M.R. Douglas and B. Fiol ``D-branes and 
Discrete Torsion II'', hep-th/9903031.}
\lref\mono{A. Klemm, W. Lerche, S. Theisen and S. Yankielowicz ``On the 
Monodromies of N=2 Supersymmetric Yang-Mills Theory'', hep-th/9412158.}
\lref\beilin{A. A. Beilinson, ``Coherent Sheaves on $\IP^n$ and Problems of
Linear Algebra'', Funct. Anal. Appl. {\bf 12} (1978), 214.}
\lref\ag{P. Argyres and M.R. Douglas, ``New Phenomena in SU(3) 
Supersymmetric Gauge Theory'', Nucl. Phys. {\bf B 448} (1995), 93.}
\lref\sv{A. Shapere and C. Vafa, ``BPS Structure of Argyres-Douglas 
Superconformal Theories'', hep-th/9910182.}

\Title{\vbox{\baselineskip12pt
\hbox{RUNHETC-2000-55}
\hbox{hep-th/0012079}
}}
{\vbox{\centerline{The BPS Spectrum of ${\cal N}=2$ SU(N) SYM}
\centerline {and Parton Branes}}}

\centerline{Bartomeu Fiol}

\bigskip
\medskip

{\vbox{\centerline{ \sl New High Energy Theory Center}
 \centerline{\sl Rutgers University}
\vskip2pt
\centerline{\sl Piscataway, NJ 08855, USA }}
\centerline{ \it fiol@physics.rutgers.edu }}

\bigskip
\bigskip
\noindent

We apply ideas that have appeared in the study of D-branes on Calabi-Yau 
compactifications to the derivation of the BPS spectrum of field theories.
In particular, we identify an orbifold point whose fractional branes
can be thought of as ``partons'' of the BPS spectrum of ${\cal N}=2$ pure
$SU(N)$ SYM. We derive the BPS spectrum and lines of marginal stability 
branes near that orbifold, and compare our results with the spectrum of
the field theories.

\noindent

\bigskip

\Date{December, 2000}

\listtoc \writetoc

\newsec {Introduction}

Over the last year, a framework for the determination of the classical 
spectrum of BPS branes of IIA string theory on Calabi-Yau varieties has been
developed, valid throughout the compactification moduli space (see \cate\ for 
the state of the art of this program and an extensive list of references). 
One of the main ingredients of
the emerging picture is that we can think of the BPS spectrum as boundstates
of a finite number of ``parton'' branes (e.g., fractional branes near an 
orbifold, $L=0$ boundary states at a Gepner point). These parton branes 
are rigid, in
the sense that they have no moduli space. Mathematically, they provide
a basis for the K theory of the Calabi-Yau. There has been much work devoted
to a better understanding of this finite set of branes for different
compactifications, and the determination of their large volume charges
\refs {\quintic, \dg, \noncompact, \iqbal, \dd, \govinda, \tomasi, \pmayr}.

Among the many applications of D-branes on Calabi-Yau manifolds, a specially
fruitful one has been the derivation of many non-trivial nonperturbative 
results of ${\cal N}=2$ quantum field theories \refs {\selfdual, \mirror}, 
building on earlier work \shamit . By suitably choosing a local 
compactification geometry and taking a decoupling limit, one can study a host 
of field theories with different gauge groups and matter content. This 
philosophy has come to be known as ``geometric engineering of quantum field 
theories'' \geoeng , and it has a number of advantages: it is 
very systematic, gives a rationale for the unexpected appearance of 
geometrical objects in the Seiberg-Witten \sw\ solution of these theories, 
and allows the study of new theories without a known Lagrangian formulation.

In the present paper we relate recent developments in the study of D-branes
on Calabi-Yau manifolds with the BPS spectrum of ${\cal N}=2$ field theories. 
The essential idea is the 
following; for concreteness, we will consider the case of pure $SU(N)$ 
SYM. Recall that along the moduli space of ${\cal N}=2$ $SU(N)$ SYM, there 
is a set of $N(N-1)$ monopoles or dyons that can go massless, and we can pick 
up $2(N-1)$ of them to form a basis of vanishing cycles \lercherev. We claim 
that these $2(N-1)$ potentially massless dyons constitute
another example of ``partons'', and the rest of the spectrum can be thought
of as boundstates of them. For instance, for $SU(2)$, the monopole and the
fundamental dyon that go massless in the strong coupling constitute such a 
set of ``partons'', and the $W^+$ and the tower of dyons present in the weak
coupling appear as boundstates of monopoles and fundamental dyons. From 
the string theory point of view, this amounts to a shift in perspective with 
respect to geometric engineering: the starting point of geometric engineering 
is given by a IIA compactification on a geometry of 2-cycles, and D2-branes 
wrapping about them, that correspond to the perturbative (electric) degrees 
of freedom of the field theory. Here, we compute the BPS spectrum in the 
non-geometric phase of the string theory compactification; in particular, we 
identify the parton branes, which correspond to D4-branes 
wrapping 4-cycles, with the basis of vanishing cycles in the field theory, so 
we obtain a description of the field theory spectrum in terms of magnetic
degrees of freedom\foot {This is always on the type IIA side. In the mirror
type IIB side, the electric magnetic duality of the theory is manifest, as 
D3 branes correspond to both electric and magnetic particles in the field 
theory.}. 

Our strategy is the following: we start by identifying an orbifold point 
$\IC^3/\IZ_{2N}$ in the moduli space of the non-compact Calabi-Yau used to 
geometrically engineer $SU(N)$. Once we have the worldvolume theory of the
branes at that orbifold, we can in principle determine the BPS spectrum of
that compactification. As explained in \refs {\pistab , \noncompact , \fm} 
near the orbifold this is a two step process: if we have a set of $k$ 
different kinds of ``partonic'' branes, first we have to determine 
for which values of $(n_1,\dots ,n_k)$ there is a vacuum 
configuration, compatible with the superpotential, breaking the original 
gauge group $U(n_1)\times \dots \times U(n_k)$ down to $U(1)$. If 
there is such a configuration for $(n_1,\dots n_k)$, then a boundstate of 
$n_1$ times the first parton, $n_2$ times the second parton and so on, can 
exist somewhere in moduli space. The second part of the procedure is to find
where in moduli this state exists. The answer depends on the Fayet-Iliopoulos 
terms and goes by the name of $\theta$-stability \refs {\king, \pistab}. 

As we will show, in a particular neighborhood of the orbifold, we are able
to identify the spectrum of BPS branes with the BPS states of the $SU(N)$ SYM
field theory. The strong coupling spectrum of these field theories was
recently derived by Lerche \lerche , by considering boundary states of a
Gepner model in the mirror Calabi-Yau.

It is clear from particular examples \noncompact , that once we move 
sufficiently away from the orbifold, it is generically not true that the
BPS spectrum can be described as boundstates of positive numbers of 
fractional branes. The reason is the following: each fractional brane has a
central charge whose phase determines which particular ${\cal N}=1$ 
supersymmetry is preserved, from the bulk ${\cal N}=2$. The crucial claim is 
that boundstates of different fractional branes can be described by a (softly
broken) ${\cal N}=1$ theory, even though each fractional brane in the 
boundstate 
may have different phase for the central charge. At the orbifold point, all 
the central charges of the fractional branes are parallel, so in the 
neighborhood of the orbifold the differences among the phases is small and 
the previous claim is justified. As we move away from the orbifold, 
eventually the phases of the central charges differ significantly, and we 
can encounter boundstates of branes that at the orbifold had antiparallel 
central charges\foot {The natural way of keeping track of this possibility 
is by considering the derived category of the category of quiver 
representations \cate . This introduces a (useful!) redundancy in the 
description, and in this sense resembles a gauge symmetry.}. In the light of
these remarks, it is not {\it a priori} obvious that an analysis near the
orbifold should suffice to recover the field theory spectrum. A better
understanding of why this is the case would require considering the periods
of these Calabi-Yau backgrounds.

The organization of the paper is the following. In section 2, after briefly
recalling the Calabi-Yau used for the geometric engineering of SU(N), we 
describe an orbifold point in the moduli space of that Calabi-Yau, and 
derive the worldvolume theory of the D-branes at that orbifold. Our next 
task is to obtain the spectrum of classical boundstates arising from that
worldvolume theory, and discuss the jumps in the spectrum that take place 
near the orbifold; that we do in 
section 3, and in section 4 we compare our results with the spectrum of SU(N)
and the known lines of marginal stability. In section 5, we state our 
conclusions.

\newsec {${\cal N}=2$ $SU(N)$ SYM and fractional branes.}

In this section we identify a non-compact orbifold, $\IC^3/\IZ_{2N}$, 
as the non-geometric phase of the local IIA Calabi-Yau compactification used 
for the geometric engineering of ${\cal N}=2$ $SU(N)$ SYM \geoeng . Once we 
have identified that 
orbifold, we can use the techniques of \dm\ to derive the ${\cal N}=1$ 
worldvolume theory for branes on that geometry. In subsequent sections, we 
will analyze the spectrum of boundstates of those theories.

\subsec {Geometric engineering of SU(N).}

It will be useful to recall the setup for the geometric engineering of 
$SU(N)$\foot {For reviews on geometric engineering, see 
\refs {\mayr,\lercherev, \klemrev}.}. The 
starting point is to consider an $A_{N-1}$ singularity in six dimensions, 
since at this singularity, type IIA develops an enhanced $SU(N)$ gauge 
symmetry. Next, we want to compactify down to four dimensions, breaking 
half of the supersymmetry on the way. To accomplish this, we fiber the 
$A_{N-1}$ singularity over a base $\IP^1$. Recall that a $A_{N-1}$ singularity
can be thought of as a $\IC^2/\IZ_{N}$ orbifold blown up by $N-1$ $\IP^1$'s.
We will denote this geometry of $N-1$ $\IP^1$'s fibered over a base $\IP^1$
by $X$.

Since we are interested in extracting field theory results from this 
compactification of string theory, we need to decouple the effects of
gravitational and massive string excitations. This amounts to taking the 
limit where the size of the base $\IP^1$ goes to $\infty$, whereas the 
sizes of the $\IP^1$'s in the fiber go to zero. This keeps finite the mass 
of the $W^\pm$ bosons, which arise from D2-branes wrapping the 2-cycles 
of the fiber $\IP^1$'s. On the other hand, the magnetically charged states 
in the field theory arise from branes 
wrapping 4-cycles, and in the limit that the volume of the base is sent to 
infinity, these states decouple from the perturbative theory.
 
Ultimately, we are interested in the vector moduli space of this 
compactification, and for this purpose it is crucial to consider type IIB on 
the mirror geometry $\hat X$. The reason for this 
is that, while for both type IIA on $X$ and type IIB on $\hat X$, the vector 
moduli space is free of quantum string corrections (as in both cases the 
dilaton sits in a hypermultiplet), on the type IIA side we would have to 
deal with world sheet instanton corrections, whereas on the type IIB side
there are no such corrections. An easy way to understand this 
is to note that the vector moduli space for type IIB encodes the size of 
the 3-cycles of $\hat X$, and neither the fundamental strings nor the IIB 
BPS D-branes can wrap a 3-cycle to produce an instanton. Therefore, a purely 
classical description of type IIB on $\hat X$ encodes all the nonperturbative 
physics of the quantum field theory.

It is convenient to describe this geometry by a two dimensional linear sigma 
model \witten. This description, or more precisely, the equivalent toric
diagram, will be specially helpful to determine the non-geometric phase.
As we just reviewed, for $SU(N)$ with no matter, we have to consider IIA on a 
$A_{N-1}$ ALE singularity fibered over a $\IP^1$. For each 2-cycle we 
introduce a U(1), 
and since we have a $\IP^1$ in the base and $N-1$ $\IP^1$'s as fibers, the 
gauge group will be $U(1)^N$. The matter content is given by $N+3$ chiral 
fields, whose charge vectors with respect to the N $U(1)$'s are

$$\vbox{
\offinterlineskip
\halign{
\strut \quad $#$ \cr
v_b=(1,1,-2,0,0,\dots,0) \cr
v_{f_1}=(0,0,1,-2,1,\dots,0) \cr
\vdots \cr
v_{f_{N-1}}=(0,0,0,\dots,1,-2,1) \cr }}$$

The geometry under consideration does not have odd cycles, and this 
translates into not having a superpotential in the linear sigma model. The 
toric diagram has vertices given by 

$$\nu _i = \left(\matrix{0&1\cr 0&-1\cr 0&0\cr 1&0\cr 
\vdots & \vdots \cr N &0 }\right) $$

\ifig\toricone{The toric diagram for SU(3).}
{\epsfxsize2.0in\epsfbox{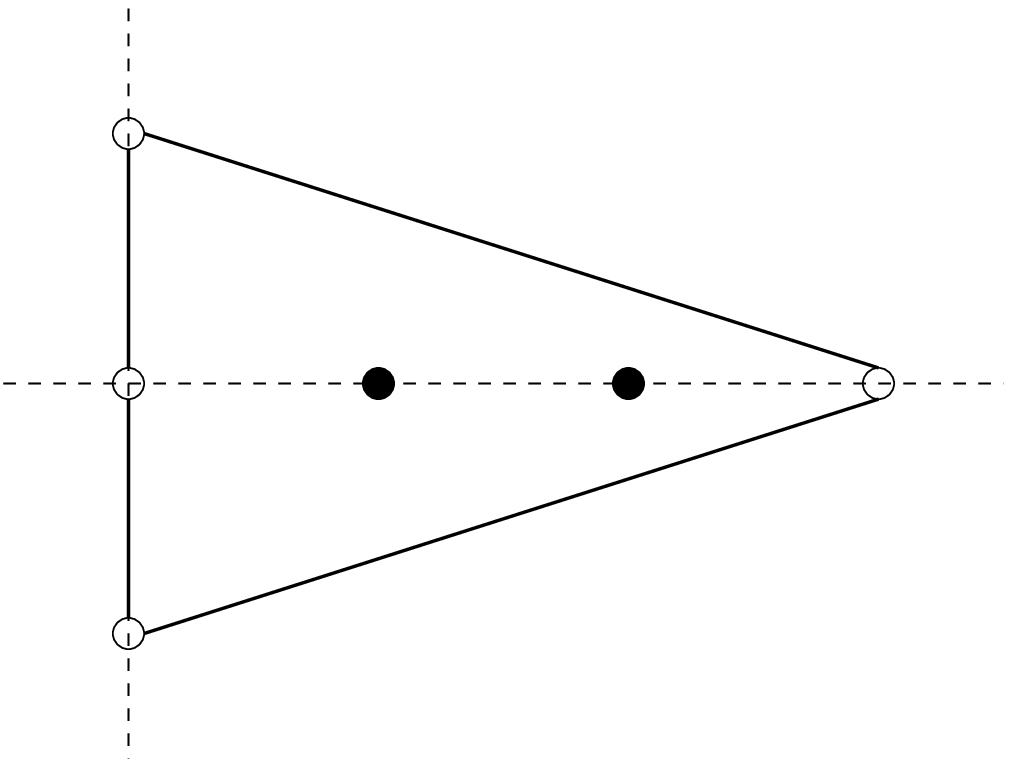}}

This is displayed in \toricone\ for $SU(3)$. The $N-1$ interior points 
correspond to the $N-1$ compact divisors. 

\subsec {The orbifold limit.}

So far, we have described the standard way to extract ${\cal N}=2$ $SU(N)$ 
SYM from a Calabi-Yau. An important ingredient is that the charged spectrum 
of the field theory appears from considering 
branes wrapping cycles of a Calabi-Yau, a topic which has received a lot
of attention lately\foot {see \dtopics\ and \dbranes\ for overviews on the 
physics and the mathematics involved, respectively.}. One of the central 
ideas in the emerging framework is that the whole spectrum of BPS 
states can be thought of as boundstates of a finite set of branes. We 
would like to determine this set of ``parton'' branes for the geometry just 
described, and relate it with the spectrum of $SU(N)$ ${\cal N}=2$ SYM. 
To do so, we need to go to a point in the moduli space of this geometry 
where we have a handle on the spectrum of BPS D-branes and their 
worldvolume theories. To accomplish this, we take a different limit that the 
one just described: keeping the $\IP^1$'s in the fiber blown down, we shrink 
the size 
of the base $\IP^1$ of $X$. Formally, in the $t_b \rightarrow -\infty$, we
reach a solvable point. The resulting non-geometic phase can be described
as follows \katz\foot {We arrived to this result by different arguments than 
those presented here. I am indebted to 
S. Katz for explaining this method to me, and for providing me with the 
lecture notes \katz\ prior to publication.}: take the vertices of the toric 
diagram in $\IZ^3$, $(0,-1,1), (0,1,1), (N,0,1)$, and equate the corresponding
monomials to $(1,1,1)$

$$(t_3^N,t_1^{-1}t_2,t_1t_2t_3)=(1,1,1)$$

the solution $t_1=t_2=\epsilon$, $t_3=\epsilon ^{-2}$, with $\epsilon ^{2N}=1$
 describes the orbifold $\IC^3/\IZ _{2N}$ with spacetime action given by

$$(z_1,z_2,z_3)\rightarrow (e^{2\pi i\over 2N}z_1, e^{2\pi i\over 2N}z_2,
e^{-2{2\pi i\over 2N}}z_3)$$

This orbifold is the point in the moduli space of the geometry $X$ 
we are going to focus on\foot {The reason we obtained an orbifold is that 
the toric diagram we started with was simplicial \katz .}. There are $2N$ 
fractional branes at this orbifold point, and they constitute what we call
the basis of parton branes for the BPS spectrum for this compactification.
To better understand the relation of
this orbifold with the geometry we started with, an $A_{N-1}$ fibration over
$\IP^1$, we can study the homology of this orbifold. If we denote the 
generator of $\IZ_{2N}$ by $g$, the action of the k$th$ element of 
$\IZ_{2N}$ is

$$g^k\ :\ (z_1,z_2,z_3)\rightarrow (e^{k{2\pi i\over 2N}}z_1,
e^{k{2\pi i\over 2N}}z_2, e^{-2k{2\pi i\over 2N}}z_3)$$

There is a complex line of $\IC^2/\IZ_2$ singularities, $(0,0,z_3)$, caused 
by $g^N$. We can compute the orbifold cohomology following \zaslow, and the 
twisted sector contribution is $h^{1,1}=N$, $h^{2,2}=N$. However, one of the
elements of $h^{1,1}$, the one coming from the $g^N$ twisted sector, is not a 
normalizable form on the resolved space, so as in \dg, we conclude that it
does not correspond to a compact 4-cycle. All told, we have $N$ 2-cycles and 
$N-1$ compact 4-cycles, which indeed matches the homology of $A_{N-1}$ 
fibered over $\IP^1$. The picture is then that for each point in $(0,0,z_3)$
we have a $\IP^1$, forming a non-compact 4-cycle $\IC \times \IP^1$, but at 
the origin $(0,0,0)$ there are extra shrunk cycles. 

What is the relation between this orbifold and the field theory?\foot { I 
would like to thank D.E. Diaconescu and C. Vafa for discussions on this 
point}. The vector moduli space of this string compactification is $N$ 
complex dimensional, whereas the moduli space of the corresponding field 
theory is $N-1$ complex dimensional, and can be regarded as a hypersurface 
in the former one. In particular, the orbifold point we just described is 
not sitting in the moduli space of the field theory, and one might worry 
that, starting at the orbifold, by the time we get to the hypersurface that 
corresponds to the field theory, the phases of the central charges have 
changed enough as to render the quiver theory approximation invalid. We should
then consider the flow of the gradings and the derived category \cate\ to
study the spectrum. A better understanding of why a particular 
neighborhood of the orbifold reproduces the expected spectrum of the field 
theory would require a full analysis of the moduli space and periods of this
string theory compactification.

Another point to take into account is that as we shrink the base $\IP^1$, we
open the possibility for D2 branes to wrap that 2-cycle, yielding new 
$W^{\pm}$ not present in the weak coupling. The appearance of this 
nonperturbative $SU(2)_{base}$ in the strong coupling limit of geometric 
engineering of $SU(N)$ was discussed in \refs {\mirror, \mayr}, building on 
earlier work \refs {\km, \kmp}. When we compare the spectrum of this string 
theory compactification with that of the $SU(N)$ field theory, we have to 
identify which fractional branes wrap the 2-cycle corresponding to the shrunk 
$\IP^1_{base}$ and discard them from our discussion. 

\subsec {The worldvolume theory.}

Finally, we are ready to derive the ${\cal N}=1$ theory of the branes at the
 orbifold. To do so, we apply the techniques introduced in \dm . Recall 
that in an orbifold $\IC^n/\Gamma$, there is a 1-to-1 correspondence between 
fractional branes and irreducible representations $r_i$ of $\Gamma$. In the 
present case we have $|\Gamma|=2N$ different fractional branes; if we want to 
consider a configuration with $n_1$ fractional branes of the first kind, $n_2$
fractional branes of the second and so on, we need to take as representation
$R=\sum _i n_i\; r_i$. The result is a ${\cal N}=1$ theory, with gauge
group $U(n_1)\times \dots \times U(n_{2N})$, $2N$ chiral fields $X_{i,i+1}$
transforming in $(n_i, \bar n_{i+1})$, $2N$ chiral fields $Y_{i,i+1}$
transforming in $(n_i, \bar n_{i+1})$, and $2N$ chiral fields $Z_{i,i-2}$
transforming in $(n_i, \bar n_{i-2})$. 

\ifig\quiverfig{The quiver diagram for the N=3 orbifold.}
{\epsfxsize2.0in\epsfbox{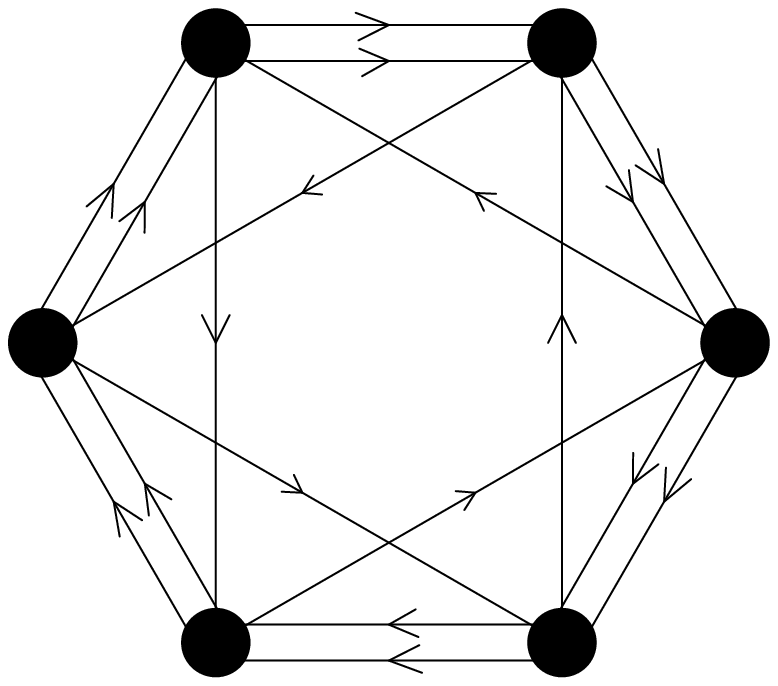}}

This is represented in \quiverfig\ for the $N=3$ orbifold, 
$\IC^3/\IZ_6$. The superpotential of the gauge theory is given by the usual 
reduction of the original ${\cal N}=4$ one, 

$${\cal W}=\hbox {tr } \sum _ {i=1}^{2N} (X_{i,i+1}Y_{i+1,i+2}-Y_{i,i+1}
X_{i+1,i+2})Z_{i+2,i}$$

which leads to the F-flatness conditions

\eqn\fflat{\eqalign{
X_{i,i+1}Y_{i+1,i+2}=Y_{i,i+1}X_{i+1,i+2} \cr 
Z_{i+2,i}X_{i,i+1}=X_{i+2,i+3}Z_{i+3,i+1} \cr
Z_{i+2,i}Y_{i,i+1}=Y_{i+2,i+3}Z_{i+3,i+1} \cr}}

In addition to the superpotential, the ${\cal N}=1$ worldvolume theory admits
Fayet-Iliopoulos terms for the $2N$ $U(1)$ factors, $\zeta_i$, 
$i=1,\dots, 2N$. These FI terms can be written in terms of the NS twist 
fields via  a discrete
Fourier transform \dgm . In the previous subsection we computed the orbifold
cohomology, or put differently, the RR ground states. The NS twist fields 
$\phi_k$ are related by supersymmetry, 
and in principle we have 2N complex NS fields, 2 coming from the $g^N$ 
twisted sector, and one from each of the remaining 2N-2 twisted sectors. 
There is a reality condition, $\phi _k=\phi _{2N-k}^*$, so finally we have 
$N$ complex NS fields. These correspond to the $N$ complexified K\"ahler 
moduli. Now the FI terms can be read from the coupling \dgm

$$\sum _k \int \; \phi _k \hbox {Tr }\gamma (g^k) D$$

where $D$ is the matrix of auxiliary fields. The outcome is that for $N$
odd, we have two relations $\sum _{k\, odd} \zeta _k= \sum _{k\, even}
\zeta _k=0$ whereas for $N$ even, we only have one relation $\sum _k\zeta _k
=0$. This comes about because for $N$ odd $\IZ_{2N}=\IZ_2\times \IZ_{N}$, but
the same is not true for even $N$. The upshot is that we have 2N-2 FI 
independent terms for $N$ odd and $2N-1$ for $N$ even. In any case, we 
conclude that the D-branes can not explore the whole of the
K\"ahler moduli space, which is $N$ complex dimensional.

What is the relation between the spectrum of this string compactification and
the spectrum of ${\cal N}=2$ $SU(N)$? To identify the fractional branes with 
states in the field theory our guide will be the intersection matrix, 
$I_{a,b}=\hbox{Tr}_{ab}\;(-1)^F$ \refs {\index, \quintic}, since in the four
dimensional field theory, when the D-branes reduce to particles, $I_{a,b}$ 
corresponds to the Dirac-Schwinger-Zwazinger (DSZ) product.

Now, since the arrows in the quiver stand for (the bosonic partners) of 
fermionic massless zero modes, one can suspect that it is possible to read
off the intersection matrix of the fractional branes of an orbifold from the
quiver. Indeed, for an abelian orbifold $\IC^n/\Gamma$ with spacetime action
$z_i\rightarrow e^{{2\pi i\over |\Gamma|}w_i}z_i$, if we denote by $g$ the 
$|\Gamma| \times |\Gamma |$ shift matrix, the intersection matrix for the
fractional branes is

$$\hbox {I}_{ab}=\Pi \; (1-g^{w_i})$$

which for a Calabi-Yau n-fold ($\sum w_i=0$ mod $n$),  is completely
symmetric or antisymmetric depending on the parity of $n$. In our case,

\eqn\intmat{I_{a,b}= (1-g)(1-g)(1-g^{-2}) = -2g+2g^{-1}-g^{-2}+g^2}

What makes this intersection matrix relevant for our discussion is that it
is exactly (minus) the intersection matrix of vanishing cycles of $SU(N)$, or 
put differently, the DSZ product for a basis of the potentially massless 
dyons of $SU(N)$, with magnetic and electric charges \lerche,

\eqn\vancy{\left (\matrix{[\alpha _1,0]\cr [-\alpha_1, \alpha _1]\cr \vdots\cr
   [\alpha _i, (i-1)\alpha _i]\cr [-\alpha _i, (2-i)\alpha _i]\cr \vdots \cr
   [\alpha _N,\sum (1-k)\alpha _k ]\cr [-\alpha _N,\sum (k-2)\alpha _k]}
\right ) }

where $\alpha _i$, $i=1,\dots ,N-1$ are the simple roots of $su(N)$ and
$\alpha _N=-\sum _i \alpha _i$. This suggests that the fractional branes
we found at the orbifold correspond, in the field theory limit, to dyons
whose magnetic charges are simple roots of the $su(N)$ algebra, and whose 
electric charges can be chosen as in \vancy. It was stablished in \hollow,
that, at least in the weak coupling, all the particles of $SU(N)$ SYM have
magnetic charge a root of the $su(N)$ algebra, and it seems quite natural 
that those whose magnetic charge is a simple root can play the role of 
partons for the rest. 

Note that for $su(N)$ we have $N-1$ positive simple roots, and indeed the 
Seiberg-Witten solution for $SU(N)$ is given in terms of a $g=N-1$ Riemann 
surface with $2(N-1)$ independent 1-cycles \lercherev. To recover 
the states with negative magnetic charge, with respect to these $2(N-1)$, 
we can choose to
add two extra 1-cycles with $\alpha_N$, as in \vancy, or stick just to a set
of independent cycles and allow for negative coefficients. This last option 
is more in the line of \noncompact, where antiparticles near the orbifold 
came from quivers representations with all the $n$'s negative. This leads us 
to identify $2(N-1)$ of the $2N$ fractional branes at the orbifold, with the 
independent vanishing cycles and the corresponding field theory particles. 
Furthermore, the quiver formed by the two adjacent nodes that we take away

\ifig\takeaw{The quiver diagram for the N=3 orbifold.}
{\epsfxsize1.0in\epsfbox{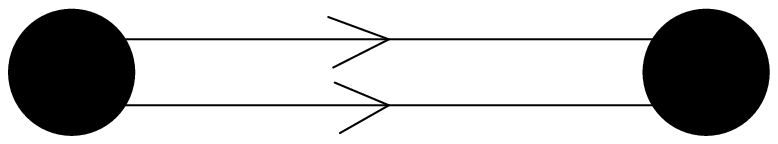}}

gives, by the Beilinson construction \beilin\ , the coherent sheaves over 
$\IP^1$. Since the stable sheaves on $\IP^1$ can be identified with the BPS 
spectrum of SU(2), it is reasonable to assume that these two 2 fractional 
branes are 
charged under the base $\IP^1$, and we should discard them in discussing the 
relation of the BPS D-brane spectrum with the $SU(N)$ spectrum. The upshot of 
this discussion is that we truncate the quiver gauge theory, cutting out two
adjacent nodes, and keeping $2N-2$ nodes.

The intersection matrix \intmat\ has also appeared recently \lerche\ in the 
study of a Gepner point in the moduli space of the type IIB mirror geometry, 
$\hat X$, to which we turn our attention next.

\subsec {The mirror picture.}

Recently, Lerche \lerche\ considered a Gepner point in the moduli space of 
the type IIB geometry $\hat X$. It is claimed in \lerche\ that that Gepner 
point corresponds to the origin of $SU(N)$ moduli space. This moduli space 
is $N-1$ complex dimensional, and at the origin there is a $\IZ_{2N}$ global 
symmetry, $u_k\rightarrow e^{2\pi i\over 2N}u_k$, where $u_k$, $k=2,\dots,N$
are the Weyl coordinates \mono. The boundary states of that
coset model are then identified with the BPS spectrum of ${\cal N}=2$ $SU(N)$
SYM at strong coupling. The role of parton branes, played on the IIA 
side by the fractional branes, is played here by the $L=0$ 
A-type rational boundary states. The starting point of \lerche\ is a LG 
potential 

$$W=x^N+{1\over z_1^{2N}}+{1\over z_2^{2N}}-\sum _{k=2}^N u_k x^{N-k} 
(z_1z_2)^{-k} $$

where $u_k$ are coordinates for the $N-1$ complex dimensional moduli space.
In particular, the point $u_k=0$ corresponds to the coset model

$$\left ( {SU(2)_{N+2}\over U(1)}\times {SL(2)_{2N+2}\over U(1)}
\times {SL(2)_{2N+2}\over U(1)}\right )/\IZ_{2N}$$

The main point of \lerche\ 
was to prove that the A-type $L=0$ rational boundary states of this Gepner 
model have the same intersection matrix \intmat\ than the basis \vancy\ of 
vanishing cycles of ${\cal N}=2$ $SU(N)$ SYM. This follows
if we grant that the factors in the intersection matrix coming from 
the different minimal models, all diagonalize in the same 
basis, something that for the ordinary minimal models happens for 
the B-type boundary states \quintic . It would be interesting to derive 
this rule from
a careful analysis of this coset model. Note also that the mirror geometry
$\hat X$ does not have any even cycles. This should translate into the fact
that this coset model does not have any B-type boundary states.

\newsec {The spectrum of BPS states near the orbifold.}

In the previous section, we derived the worldvolume theory for a 
configuration of  $(n_1,\dots,n_{2N})$ fractional branes. The next question 
we would like to ask is in which cases they form a BPS boundstate, and how the
answer may change as we move in moduli space. The general procedure was 
introduced in \refs {\pistab, \noncompact}, and explained in detail in \fm, 
so here we will be quite brief. 

First, the criterion for having a boundstate is that the vevs of the chiral 
fields break the original gauge group completely, except for the diagonal 
U(1), which is always present for these theories, and will represent the 
center of mass motion. 

The next thing that we require to the boundstates is that they are BPS. Away 
from the orbifold, the general configuration of different fractional branes 
will break all supersymmetry, as each preserves a different ${\cal N}=1$ 
subalgebra of the original ${\cal N}=2$. The claim is that this supersymmetry 
breaking is quite a mild one, caused by a constant non-zero potential coming 
entirely from D terms. In the language of $\theta$-stability \refs {\king,
\pistab} this means that for a given set of values $(n_1,\dots , n_{k})$, we 
look for $\theta$-stable configurations with the components of $\vec \theta$ 
related to the physical FI terms $\zeta $ by

\eqn\thetzet{\theta _i=\zeta _i-{\vec n\cdot \vec \zeta \over \vec n\cdot e}}

where $e=(1,\dots, 1)$. The difference between $\zeta _i$ and $\theta _i$
gives precisely the constant shift in the potential just discussed.

The last ingredient in the picture is how the spectrum changes as we move
in K\"ahler moduli space. The field theory counterpart would be the 
determination of the
lines of marginal stability in the quantum moduli space. The answer depends
entirely on the D terms, as the holomorphic properties of the states, dictated
by the superpotential of the worldvolume theory, are independent of K\"ahler
moduli. More concretely, near the orbifold, $\theta$-stability depends
explicitly on K\"ahler moduli, through the FI terms. A criterion for 
stability based on the periods of the Calabi-Yau, and therefore exact in
$\alpha '$, was presented in \pistab . Note that, 
as in geometric engineering, we need of mirror symmetry to provide the exact 
periods if we want a complete discussion of the lines of marginal stability.

These ideas have a nice mathematical counterpart 
near the orbifold, known as quiver theory, summarized in the following
table

$$\vbox{
\offinterlineskip
\halign{

\strut \vrule \quad $#$ & \vrule \quad $#$ \vrule \cr
\noalign{\hrule}
\hbox {Worldvolume theory} & \hbox {Quiver theory}\cr
\noalign{\hrule}
\hbox {F-flatness conditions} & \hbox {Quiver with relations} \cr
\hbox {Single boundstate} & \hbox {Schur representation} \cr
\hbox {``Quasi'' susy vacuum} & \hbox {$\theta$-stable representation} \cr
\noalign{\hrule}
}}$$

Let's briefly recall how to obtain the spectrum of boundstates. For more
details, see the appendix of \noncompact . For configurations with $\vec n
=(n_1,\dots , n_k)$ fractional branes and where the F-flatness conditions 
are trivially satisfied (by setting to zero some vev's), the expected 
dimension of the moduli space is 

\eqn\expdim{d(\vec n)=1-{1\over 2}\vec n ^T\cdot C\cdot \vec n}

with $C$ the generalized Cartan matrix associated with the quiver. This is 
quite easy to understand; we count the number of parameters in the
gauge group and subtract the number of entries in the matrices representing
the vevs of the chiral fields. We call imaginary and real roots those $\vec n$
for which $d(\vec n)\geq 1$ and $d(\vec n)=0$, respectively. A Schur root
is an imaginary or a real root, but the opposite is not true, so after finding
all imaginary and real roots, we have to determine which ones of those are
Schur roots. We are not aware of a systematic procedure, but for our problem,
it will be fairly easy to decide in many cases.

\subsec {Range of validity.}

Before we embark in the study of the spectrum of classical ($g_s=0$)
boundstates of this compactification, we would like to discuss the range of 
validity of the quiver theory and of $\theta$-stability. 

As mentioned in the introduction, a claim central to recent work on BPS 
D-branes on Calabi-Yau manifolds \refs {\pistab, \noncompact, \dd, \cate} 
is that these BPS states can be described by ${\cal N}=1$ theories, even
though the constituent parton branes preserve different ${\cal N}=1$
supersymmetries. At an orbifold point typically the central charges are 
real (the only contribution coming from the B field at the singularity), so
they are aligned. As we move away from the orbifold, the central charges of
the different fractional branes will be no longer aligned, and eventually it
can happen that two fractional branes $A$ and $B$, that near the orbifold 
had almost parallel central charges, now have them almost antiparallel, so 
at that point in moduli space the possible boundstate is between $A$ and 
$\bar B$, the antibrane of $B$. This is discussed in detail in \cate. By 
the time we reach this point, the quiver theory description has broken down.

On the other hand, $\theta$-stability is only valid at linear order in the 
FI terms, so it breaks down as soon as the periods are no longer linear in 
the FI terms. 

For the present case, we will show that the field theory particles 
correspond to BPS states in the string theory that can be described with
quiver theory, but $\theta$-stability only gives the lines of marginal
stability somewhere near orbifold, and a complete stability
analysis would involve the full $\Pi$-stability condition.

\subsec {The boundstates of the worldvolume theory.}

The truncated quiver with $2N-2$ nodes can be pictured more conveniently as

\eqn\truncquiver{
\diagram
& V_1 & \pile {\rTo \\ \rTo} & V_2 \\
& \uTo  & \pile {\ldTo \\ \ldTo} & \uTo \\
& V_3 & \pile {\rTo \\ \rTo} & V_4 \\
& \uTo  & \pile {\ldTo \\ \ldTo} & \uTo \\
& V_5 & \pile {\rTo \\ \rTo} & V_6 \\
\enddiagram}

where the $V_i$ are vector spaces for representations of the quiver.
As we will see, it turns out that all the states of the $SU(N)$ SYM
theory, can be identified with representations of the previous quiver with 
the diagonal arrows set to zero,

\eqn\swquiver{
\diagram
& V_1 & \pile {\rTo \\ \rTo} & V_2 \\
& \uTo  &  & \uTo \\
& V_3 & \pile {\rTo \\ \rTo} & V_4 \\
& \uTo  &  & \uTo \\
& V_5 & \pile {\rTo \\ \rTo} & V_6 \\
\enddiagram}

There are definitely more boundstates in the former quiver,
meaning that there are BPS states in the string compactification that don't
appear in the $SU(N)$ spectrum. We will see that it is possible to choose the
FI terms in such a way that there is a region near the orbifold where these
extra states are not present.

First, we are going to study the spectrum of boundstates when only the 
vertical arrows have non-zero vev,

\eqn\backtwo{
\diagram
& V_{2k-1} & \rTo & V_{2k-3} & \rTo & \dots & \rTo & V_{1} \\
\enddiagram}

The F-flatness conditions are trivially satisfied in this case. The expected
dimension of the moduli space of the gauge theory is

$$d=1-\left (n_1^2+n_3^2+\dots +n_{2k-1}^2-n_1n_3-\dots 
-n_{2k-3}n_{2k-1}\right )$$

So the imaginary roots should satisfy

$$n_1^2+(n_1-n_3)^2+\dots +(n_{2k-3}-n_{2k-1})^2+n_{2k-1}^2 \leq 0$$

and the real roots

$$n_1^2+(n_1-n_3)^2+\dots +(n_{2k-3}-n_{2k-1})^2+n_{2k-1}^2 = 2$$

We immediately see that there are no non-trivial imaginary roots. For the 
real roots, the only possibility is that two summands are 1 and the rest 0. 
A moment's thought shows that all the solutions consist of a chain of 
adjacent $n_i=1$ and the rest of the $n_j$'s set to zero. For instance, if 
we have $n_1=n_3=n_5=1$, it describes a boundstate with one fractional brane 
of the first kind, one of the third and one of the fifth. Furthermore, these
representations break the gauge group to the diagonal $U(1)$: we start with
a gauge group $U(1)\times U(1)\times \dots \times U(1)$, and each nonzero
vev breaks the two U(1)'s it transforms under to their diagonal U(1). 
Therefore, they correspond to boundstates. In the next subsection, we will 
identify these boundstates with the potentially massless dyons.

Next we consider configurations with only the vevs of a pair of horizontal 
arrows not zero,

\eqn\kronecker{
\diagram
& V_1 & \pile{\rTo \\ \rTo} & V_2 \\ 
\enddiagram}

This is known in the math literature as the Kronecker quiver. The 
superpotential plays no role and the spectrum of boundstates is actually 
well known \scho (see also the appendix of \noncompact). The expected 
dimension of the moduli space is

$$d=1-(n_1^2+n_2^2-2n_1n_2)=1-(n_1-n_2)^2$$

so the imaginary Schur roots must satisfy 

$$(n_1-n_2)^2\leq 0\Rightarrow n_1=n_2$$

and actually only $(1,1)$ is a Schur root \kac . In the next section we will
identify it as a $W^+$ boson with electric charge a simple root of 
$su(N)$. All the real roots of this quiver are Schur \scho\ , and 
they are given by

$$(n_1-n_2)^2 =1 \Rightarrow n_1=n_2\pm 1$$

Later we will identify these states as the familiar towers of dyons with 
fixed magnetic charge, when the magnetic charge is a simple root of the 
algebra.

Now we consider quiver representations with both the horizontal and
the vertical representations turned on. The F-flatness conditions are
no longer satisfied automatically, so we are in the realm of quivers with
relations, and the methods we have been using no longer apply. Nevertheless,
we will be able to display representations that satisfy F-flatness and break 
the gauge group to U(1), corresponding to the expected positive charged gauge
fields and tower of dyons. We believe that those are the only Schur 
representations of this quiver compatible with the superpotential, but we 
don't have a proof of this claim. Consider for concreteness states with
magnetic charge given by $\alpha _1+\alpha _2$. They correspond to 
representations of

\eqn\mixquiver{
\diagram
& \IC^{n_1} & \pile {\rTo^{X_1} \\ \rTo_{Y_1}} & \IC^{n_2} \\
& \uTo_{Z_1} & & \uTo^{Z_2} \\
& \IC^{n_1} & \pile {\rTo^{X_2} \\ \rTo_{Y_2}} & \IC^{n_2}\\
\enddiagram}

the F-flatness conditions \fflat\ reduce to

$$Z_1X_1=X_2Z_2\hskip2cm Z_1Y_1=Y_2Z_2$$

If we take $n_1=n_2=1$, and nonzero vevs, $Z_1=Z_2, X_1=X_2,Y_1=Y_2$ we
satisfy the F-flatness conditions and break the gauge group to the diagonal
U(1). These kind of representations correspond to $W^+$ bosons whose electric
charge is a positive, but not simple, root of the algebra. We can describe 
another solution: take $n_1=n, n_2=n+1$ and $X_1=X_2$ and $Y_1=Y_2$ to be the 
Schur representation of the Kronecker quiver for $(n,n+1)$, $Z_1=\II _n$
and $Z_2=\II _{n+1}$. It is clear that the F-flatness conditions are 
satisfied, and it also easy to see that this choice of vevs breaks the gauge 
group to the diagonal $U(1)$: originally we had $U(n)\times U(n+1)\times 
U(n)\times U(n+1)$. By construction $X_1,Y_1$ break the first $U(n)\times 
U(n+1)$ to its diagonal U(1), and $X_2,Y_2$ do the same for the second 
$U(n)\times U(n+1)$. Finally, both $Z_1$ or $Z_2$ break the $U(1)\times 
U(1)$ we had so far to the diagonal $U(1)$. Therefore we have a boundstate. 
We will identify them in the field theory with dyons whose magnetic charge
is a positive but not simple root of the algebra. Note that the presence of 
the F-flatness conditions is crucial to avoid the presence of many unwanted 
states: we could
have considered a $(n,n+1,m,m+1)$ representation, with the Schur 
representations of the Kronecker quiver for the $(n,n+1)$ and $(m,m+1)$ and
nonzero matrices $Z_1,Z_2$. This would break the gauge group to U(1), but
in general it does not satisfy the F-flatness conditions. 

Finally, we can consider turning on the diagonal arrows of \truncquiver .
 As mentioned,
states with non zero vevs of these fields don't appear in the $SU(N)$ 
spectrum, but we don't have a {\it a priori} reason to discard them from 
our study. It is immediate that there are new states. For instance, we can 
turn just a pair of diagonal arrows in \truncquiver ,

$$
\diagram
& V_2 & \pile{\rTo \\ \rTo} & V_3 \\ 
\enddiagram
$$

and this is just a Kronecker quiver \kronecker\ , which has infinite
boundstates. More than that, turning on now just horizontal and diagonal
vevs

\eqn\fordone{
\diagram
& V_1 & \pile{\rTo \\ \rTo} & V_2 & \pile{\rTo \\ \rTo} & \dots &
\pile{\rTo \\ \rTo} & V_k \\ 
\enddiagram}

there is always a solution satisfying the superpotential constraints, given 
by setting all the $n_i=1$. We believe it is the only solution, but we don't
have a proof ot this claim.

\subsec {Subrepresentations and domains of stability.}

In the previous subsection, we have described the possible boundstates of
branes we have near the orbifold. To decide where in moduli space each of 
them is present, we have to check where are they $\theta$-stable. We present
the computation for a number of examples. Some of the 
novel notions of homological algebra that enter the generic picture of 
\refs {\pistab, \cate} are quite easy to understand in this limit.

We will look for subrepresentations and study the domains of stability. Let's
start with the boundstates with only vertical arrows. As we just argued,
all the non trivial
vector spaces at the nodes of the representations have dimension 1, so
we will represent them by $\IC$. For the sake of concreteness, let's consider
a boundstate of 3 fractional branes. Our considerations generalize trivially.
There are in this case two subrepresentations

$$
\diagram
& \IC  & \rTo^{\scriptstyle \simeq} & \IC  & \rTo^{\scriptstyle \simeq}
& \IC   \\ & \uDashto_{\scriptstyle 0} &      & \uDashto_{\scriptstyle 0} &
 & \uDashto_{\scriptstyle \simeq}  \\  &  0   & \rTo^{\scriptstyle 0} & 0  &
\rTo^{\scriptstyle 0} &  \IC  \\
\enddiagram
$$

and

$$
\diagram 
& \IC  & \rTo^{\scriptstyle \simeq} & \IC  & \rTo^{\scriptstyle \simeq}  
& \IC   \\ & \uDashto_{\scriptstyle 0} &      & \uDashto_{\scriptstyle 
\simeq} &  & \uDashto_{\scriptstyle \simeq}  \\  &  0   
& \rTo^{\scriptstyle 0} & \IC  & \rTo^{\scriptstyle \simeq} &  \IC  \\ 
\enddiagram
$$
 
The notation of these diagrams was introduced in \fm . The top row is the
original representation, and the bottom one is the subrepresentation. When 
the vev of a chiral field is non-zero, we perform a complex gauge 
transformation to set it to the identity map, denoted by $\simeq$. Vevs not
turned on are represented by the 0 map. By definition, there must be
an injective map from the subrepresentation to the original representation,
denoted here by the dashed vertical arrows. The commutativity of these 
diagrams is evident. In general, for a boundstate given by a chain of $k$ 
vector spaces $\IC$ and $k-1$ identity maps among them, there will be $k-1$ 
subrepresentations, being embedded in the original representation 
``by its end'', to ensure the commutativity of the diagram.

We can study now the domain of stability of these representations. Again for
concreteness, we focus in the particular example with 3 nodes. We introduce
a vector $(\theta _1,\theta _2, \theta _3)$, which must satisfy $n\cdot 
\theta =0$, which in this case reduces to $\theta _1+\theta _2 +\theta _3=0$.
Stability against decay triggered by the first subrepresentation requires
$\theta _3 >0$, while the second one requires $\theta _2+\theta _3>0$ or 
equivalently, $\theta _1 <0$. In the $\theta$-plane we have then two lines
of marginal stability for this boundstate.

Let's move now to the boundstates of the Kronecker quiver \kronecker. First 
the imaginary root (the $W^+$ boson) has a single subobject

$$
\diagram
& \IC & \pile{\rTo^{\scriptstyle \simeq} \\ \rTo_{\scriptstyle \simeq}} & 
\IC \\ & \uDashto_{\scriptstyle 0} & & \uDashto_{\scriptstyle \simeq} \\
& 0 &\pile{\rTo^{\scriptstyle 0} \\ \rTo_{\scriptstyle 0}}& \IC \\
\enddiagram
$$

This representation is then stable for $\theta _2>0$. Next
we should consider the subrepresentations of the real roots of this quiver.
Displaying them would require some work, as now we are dealing with 
vectorspaces of arbitrary dimensions. 

$$
\diagram
& \IC ^n & \pile {\rTo \\ \rTo }  & \IC^{n\pm 1} \\
\enddiagram
$$

Fortunately, if we just want to know the lines of marginal stability, we 
don't need that much. For quivers without relations, there is a theorem 
characterizing Schur roots, due to Schofield \schotwo, which will be quite 
useful for us. The generic theorem is explained in the appendix of 
\noncompact\ , and in the present case it boils down to saying that 
$(n_1,n_2)$ is 
a Schur root iff all its subrepresentations $(n_1',n_2')$ satisfy 
$n_1'/n_2' < n_1/n_2$. Now consider a particular Schur root $(n_1,n_2)$, and 
introduce a vector $(\theta_1, \theta _2)$ such that 
$n_1\theta_1+n_2\theta_2=0$. Then $(n_1,n_2)$ is $\theta$-stable if 
$n_1'\theta_1+n_2'\theta_2>0$, i.e., if $\theta _2>0$.

Finally, we will consider an example of the domain of stability for 
boundstates of \swquiver, when both horizontal and the vertical arrows are 
nonzero. For states $(n,n+1,n,n+1)$ we can have subrepresentations 
$(m_1,m_2,m_1,m_2)$ such that $(m_1,m_2)$ is a subroot of $(n,n+1)$; in 
particular $m_1/m_2 <n/n+1$

$$
\diagram
& & & \IC^n & & \pile {\rTo \\ \rTo } &  & \IC^{n+1} \\
& &\ruTo &\uDashto & & &\ruTo & \uDashto\\
&\IC^n &\pile {\rTo \\ \rTo }&\HonV[=] & & \IC^{n+1} & & \\
&\uDashto& &\uDashto & &\uDashto  & &\\
& & & \IC^{m_1}&\pile {\rTo \\ \rTo }&\VonH & \pile {\rTo \\ \rTo } 
& \IC^{m_2} \\
& &\ruTo & & & &\ruTo & \\
&\IC^{m_1} &\pile {\rTo \\ \rTo }& & & \IC^{m_2} & & \\
\enddiagram
$$

We choose a vector $\vec \theta _i$ such that 
$n(\theta_1+\theta_3)+(n+1)(\theta_2+\theta_4)=0$. This representation is
$\theta$-stable against $(m,m+1,m,m+1)$ if 
$m_1(\theta_1+\theta_3)+m_2(\theta_2+\theta_4)>0$. Using $m_1/m_2 <n/n+1$,
we see that we must require $\theta_2+\theta_4>0$. There is another possible
subrepresentation that can trigger a decay,

$$
\diagram
& & & \IC^n & & \pile {\rTo \\ \rTo } &  & \IC^{n+1} \\
& &\ruTo &\uDashto & & &\ruTo & \uDashto\\
&\IC^n &\pile {\rTo \\ \rTo }&\HonV[=] & & \IC^{n+1} & & \\
&\uDashto& &\uDashto & &\uDashto  & &\\
& & & \IC^{n}&\pile {\rTo \\ \rTo }&\VonH & \pile {\rTo \\ \rTo } 
& \IC^{n+1} \\
& &\ruTo & & & &\ruTo & \\
&0 &\pile {\rTo \\ \rTo }& & & 0 & & \\
\enddiagram
$$

Note that we can't place the 0's in the other two nodes, for the diagram
would not commute. Physically this means that among the decay products,
one of them (but not the rest) are triggering the decay \pistab . This
subrepresentation imposes $n\theta _1+(n+1)\theta _2>0$ for stability of
the original representation.

\newsec {Comparison with the ${\cal N}=2$ $SU(N)$ SYM spectrum.}

In the previous section, we have derived the boundstates of fractional
branes, and we studied the jumps in the spectrum near the orbifold, by 
resorting to a simple linear analysis. We would like now to compare
with the results for field theory. This involves a number of issues. 

i) As already mentioned, for $SU(N)$ we consider the truncated quiver
with $2(N-1)$ nodes. The antiparticles are to be thought of as representations
with all the $n$'s negative.

ii) We are going to show that a particular choice of FI terms, reproduces
the strong coupling spectrum: only the potentially vanishing states. Note
that there are $N(N-1)$ of those, not just the $2(N-1)$ corresponding to the
basis of vanishing cycles. More than that, varying the FI terms we will find
new states that also have a counterpart in the field theory. A better 
understanding of why this particular neighborhood of the orbifold reproduces
the expected spectrum of the field theory would require a full analysis of
the moduli space and periods of this string compactification.

\subsec {SU(2)}

In this case, we have two possible partons in the theory, and according 
to the identification we proposed in section 2, they correspond to
the monopole $[1,0]$ and the fundamental dyon $[-1,1]$ that go massless in 
the Seiberg-Witten solution \sw. This amounts to assign to the boundstate 
$(n_1,n_2)=n_1[1,0]+n_2[-1,1]=[n_1-n_2,n_2]$ magnetic and electric charges 
given by\foot {Our notation is as follows, $(n_1,n_2)$ represents a
boundstate of $n_1$ monopoles and $n_2$ fundamental dyons; $[q_m,q_e]$ 
represents a state of magnetic and electric charges $q_m$ and $q_e$.}

$$q_m=n_1-n_2\hskip2cm q_e=n_2$$

The worldvolume theory corresponds to the Kronecker quiver and the spectrum 
is the following: the imaginary root $(1,1)$; its charges are $[0,1]$, so
we identify it with the $W^+$ boson. Notice that the mathematical statement
that there are no $(k,k)$ boundstates, even though they satisfy the dimension
formula, corresponds to the statement that there are no particles with
charge $[0,k]$ in $SU(2)$ SYM. Next we have the real roots 
$(n_1,n_1\pm 1)$. Their charges are $[\pm 1, n_1]$, and we recognize them as
the tower of dyons with one unit of magnetic charge.

What can we say about lines of marginal stability in this case? The physical
moduli space, in the linear approximation applied in this paper, has as 
coordinates the two FI terms of the worldvolume gauge theory, $\zeta _1, 
\zeta_2$. For each $(n_1,n_2)$ we introduce a vector $(\theta _1, \theta_2)$.
In the previous section, we have performed the $\theta$-stability analysis 
for these boundstates. The result was that all of them decay when 
$\theta_2>0$. Now the relation \thetzet\ between the physical FI terms 
$\zeta_i$ and the $\theta _i$ reads in this case

$$\theta _1= {n_2\over n_1+n_2}(\zeta_1-\zeta_2) \hskip2cm
\theta _2= {n_1\over n_1+n_2}(\zeta_2-\zeta_1)$$

So the condition $\theta _2>0$ for the different boundstates $(n,n\pm 1)$
translates into a common condition in term of the FI terms, 
$\zeta_2>\zeta_1$, even though the map between $\theta$'s and $\zeta$'s 
changes for different boundstates. In other words, the linear analysis 
predicts that all these states decay at the same line of marginal stability. 
This is precisely what happens for the tower of dyons of $SU(2)$! \bife . 
This result has also been derived within the framework of geometric 
engineering \selfdual .

\ifig\loms{Line of marginal stability for the N=2 orbifold.}
{\epsfxsize2.0in\epsfbox{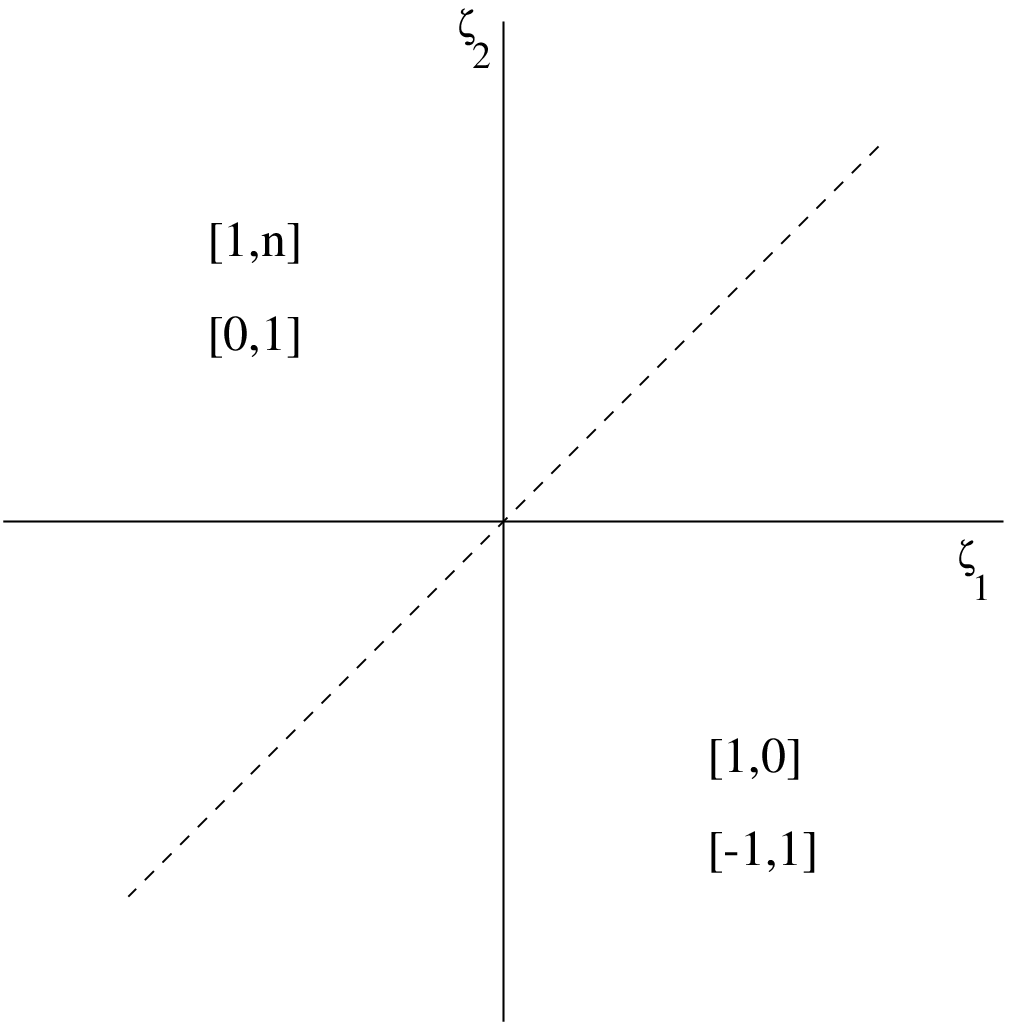}}

\subsec {SU(3)}

Already for $SU(3)$, we are not aware of a detailed description of all the 
lines
of marginal stability. A qualitative new feature is the presence in the
moduli space of points, Argyres-Douglas points \ag, where mutually non-local 
particles go massless. The spectrum of BPS states near these points was 
studied in \sv.  

The truncated quiver has 4 nodes, 

\eqn\suthree{
\diagram
& V_1 & \pile {\rTo \\ \rTo} & V_2 \\
& \uTo  & \pile {\ldTo \\ \ldTo} & \uTo \\
& V_3 & \pile {\rTo \\ \rTo} & V_4 \\
\enddiagram}

Let's start listing the possible states: first we have the four fractional 
branes $(1,0,0,0),(0,1,0,0),(0,0,1,0)$ and $(0,0,0,1)$. They are identified
with the basis of vanishing cycles. The remaining two potentially massless
dyons come from states with just the vertical arrows turned on: $(1,0,1,0)$ 
and $(0,1,0,1)$. This is the set of $N(N-1)=3\cdot 2=6$ potentially massless
dyons.

The positively charged gauge bosons are also easily identified: they 
correspond to bound states with the horizontal arrows turned on :$(1,1,0,0)$
and $(0,0,1,1)$ are the ones with electric charge $\alpha _1$ and $\alpha_2$, 
and $(1,1,1,1)$ is the one with electric charge $\alpha_1+\alpha_2$.

The towers of dyons with magnetic charges $\alpha _1$ and $\alpha _2$ 
are bound states of the Kronecker quivers: $(n,n\pm 1,0,0)$ and 
$(0,0,n,n\pm 1)$. For the positive non-simple root $\alpha_1+\alpha_2$ we 
have states described by \mixquiver\ , $(n,n\pm 1, n, n\pm 1)$. Finally, we 
have states that are not expected in the field theory: $(1,1,1,0)$, 
$(0,1,1,1)$, $(1,1,1,1)$, $(0,n,n\pm 1,0)$.

Let's describe now the lines of marginal stability for the different states.
Using the results we obtained in the previous section, we see that
the two potentially massless dyons $(1,0,1,0)$ and $(0,1,0,1)$ are 
present in the spectrum as long as $\zeta_1>\zeta_3$ and 
$\zeta_2>\zeta_4$, respectively. The $W^+$ bosons with electric charges 
$\alpha_1$ and $\alpha_2$ exist as long as $\zeta_2>\zeta_1$ and 
$\zeta_4>\zeta_3$, respectively. The $W^+$ boson with electric charge
$\alpha_1+\alpha_2$ requires $\zeta_1+\zeta_2>\zeta_3+\zeta_4$ and 
$\zeta_2+\zeta_4>\zeta_1+\zeta_3$.

The towers of dyons with magnetic charge $\alpha_1$ and $\alpha_2$ are
stable as long as $\zeta_2>\zeta_1$ and $\zeta_4>\zeta_3$. The dyons
with magnetic charge $\alpha_1+\alpha_2$ require $\zeta_2+\zeta_4>
\zeta_1+\zeta_3$ and $n(\zeta_1-\zeta_3)+(n\pm 1)(\zeta_2-\zeta_4)>0$.

Finally the states that are not present in field theory, $(1,1,1,0)$,
$(0,n,n\pm 1,0)$ and $(0,1,1,1)$ will appear when 
$3\zeta _3>\zeta_1+\zeta_2+\zeta_3>3\zeta_1$, $\zeta_3>\zeta_2$ and 
$3\zeta _4>\zeta_2+\zeta_3+\zeta_4>3\zeta_2$, respectively.

We see then that if we consider the region with $\zeta _1>\zeta_3$, 
$\zeta _2>\zeta_3$ and $\zeta_2>\zeta_4$, we always have the potentially
massless dyons present in the spectrum and none of the states that don't
appear in the field theory.

\subsec {SU(N).}

For generic $SU(N)$, we have first the boundstates consisting of only 
``back two'' chiral fields (vertical arrows in \swquiver). Let's count how 
many of them we have. First,
we have the 2(N-1) nodes. For each node except the last two, we have a 
boundstate of just two fractional branes, i.e. involving a single arrow; there
are 2(N-2) of those. Furthermore, there are 2(N-4) boundstates of 3 fractional
branes, involving two arrows, and so on. All in all, there are N(N-1) such
boundstates. From the identification of the 2(N-1) fractional branes
with the basis of vanishing cycles, it follows that these N(N-1) boundstates 
are the spectrum of potentially massless dyons. According to \lerche, this
is the strong coupling spectrum of $SU(N)$ SYM. The correspondence of our 
boundstates with the rational A-type boundary states of the Gepner model
of \lerche\ is quite clear: the nodes of the quiver are the $L=0$ boundary 
states. The boundstates with a single arrow correspond to the $L=1$ boundary 
states, the ones with two arrows are the $L=2$ boundary states, and so on.

A very similar counting, but now of representations with horizontal arrows 
turned on, yields $N(N-1)/2$ postively charged gauge bosons. Note that this
analysis can't recover the neutral gauge bosons of the field theory as they
don't arise from branes wrapping cycles.

On top of these states, we also have as potential states in the spectrum all
the boundstates with the ``forward one'' arrows. When there are only two 
nodes, they give the tower of dyons for the different simple roots of the
algebra. When we have more than one pair of horizontal arrows, we obtain
the tower of dyons for positive non-simple root. The analysis of the domains 
of stability of the different states, could be carried out as for $SU(3)$. In
particular, if we take our FI terms satisfying $\zeta _i>\zeta _{i+1}$, the
only states present in that region are the potentially massless dyons, so 
in this negihboorhod of the orbifold the spectrum coincides with the expected
BPS spectrum of the field theory.

\newsec {Conclusions.}

In this paper we have related some of the ideas that have been recently 
brought up in the study of BPS branes on Calabi-Yau varieties to the more 
familiar setting of ${\cal N}=2$ field theories. To do so we started with the
Calabi-Yau geometry used to geometrically engineer pure $SU(N)$ SYM, and 
considered the non-geometric phase, an orbifold. The advantage of studying 
this phase is that it is then very easy to obtain a set of branes that 
constitute a basis for the K-theory of the Calabi-Yau, 
namely the fractional branes at the orbifold. These fractional branes
are identified with dyons of the field theory whose magnetic charge
is a simple root of the algebra. The whole spectrum can be thought of as 
boundstates of a finite number of these states. We have displayed these 
boundstates, and performed a study of their domains of existence near the
orbifold.

As already mentioned, a crucial step in our derivation of the orbifold
point was that the toric diagram was simplicial. As this is not the case
for general ${\cal N}=2$ theories with matter, it is not straightforward
to generalize the kind of analysis we have performed here. The blowdown limit
can still be derived using the methods of \katz, but it won't be an orbifold.
On the mirror side, there is a proposed Gepner model for $SU(N_c)$ with 
$N_f=N_c-1$ flavors \lerch, so one expects this case to present some 
simplification on the type IIA side also. 

Finally, on a more general level, one can ask what are the structures behind 
the spectrum of BPS states, both in string theory compactifications and in 
${\cal N}=2$ field theories. On one hand, we can consider the derived 
category \cate , which is 
manifestly independent of vector moduli space. Another possibility is to
consider the algebra of BPS states: a universal property of the spectrum of 
BPS states for any theory is that they form an algebra \hm, which depends on 
vector moduli space. In \hm, the definition for the product of that algebra 
was given in terms of a scattering process, and obvious phase space 
considerations force an analytic continuation
to complex momenta. In \fm\ , a slightly different interpretation of the
algebra of BPS states was presented : if the coefficient $c_{ij}^k$ in the 
algebra is not zero, we say that $\phi _i$ and $\phi _j$ can form a 
boundstate $\phi _k$, with $\phi _i$ being a subobject of $\phi _k$. Notice 
that the definition is not symmetric in $ij$.

In \fm\ the notion of algebra of BPS states was reformulated near 
orbifold points. The results presented here could be then used to 
study the algebra of BPS states for field theories. Take for instance 
${\cal N}=2$ $SU(2)$ SYM. Its spectrum is given by the Kronecker quiver of 
 \kronecker , whose Schur roots correspond to the stable sheaves of $\IP^1$. 
Indeed, if we identify the rank of a sheaf in $\IP^1$ with magnetic charge 
and the first Chern number with the electric charge, we see that the tower 
of dyons correspond to line bundles ${\cal O}(k)$ on $\IP^1$, and the $W^+$ 
corresponds to the skyscraper sheaf of length one, ${\cal O}_{\IP}$. We 
have the exact sequence,

$$0\rightarrow {\cal O}(n)\rightarrow {\cal O}(n+1)\rightarrow {\cal O}_\IP
\rightarrow 0$$

This can be read as saying that a $[1,n]$ dyon and a $W^+$ boson can form
a $[1,n+1]$ dyon, with the $[1,n]$ dyon being a subobject, but not the 
$W^+$ boson. Notice that we can not have the reverse exact sequence,
as we can't have an injective map from torsion sheaves to torsion free
sheaves. This means \refs {\hm, \fm} that the structure constants 
$c^{n+1}_{n,W}\neq 0$ and $c^{n+1}_{W,n}=0$. It would be very interesting to 
determine the algebra beyond this homologic approximation, and elucidate its 
dependence on the coupling constant.

\bigskip

{\bf Acknowledgements}: It is a pleasure to thank M.R. Douglas for comments
and discussions. I have also benefited from discussions and correspondence
with P. Aspinwall, D.-E. Diaconescu, S. Katz, W. Lerche, P. Mayr, 
M. Mari\~no, C. Vafa and E. Witten. I would like to thank the Center for 
Geometry and Theoretical Physics at Duke University, and the High Energy 
groups at CalTech and Harvard, for hospitality while this paper was being
written up.

The diagrams in this paper were prepared with the package 
diagrams.tex, created by Paul Taylor. I would like to 
thank him for making it available to the scientific community 
with no charge.  

\listrefs
\end